\documentclass[fleqn,twoside]{article}
\usepackage{espcrc2}

\usepackage{epsf}
\newcommand{\AmS}{{\protect\the\textfont2
  A\kern-.1667em\lower.5ex\hbox{M}\kern-.125emS}}
\def\lsim{\raise0.3ex\hbox{$<$\kern-0.75em\raise-1.1ex\hbox{$\sim$}}}
\def\gsim{\raise0.3ex\hbox{$>$\kern-0.75em\raise-1.1ex\hbox{$\sim$}}}

\title{
\vspace*{-30pt}
{\normalsize \hfill {\sf UTHEP-482}} \\
{\normalsize \hfill {\sf UTCCP-P-146}} \\
{\normalsize \hfill {\sf October 2003}} \\
\vspace*{15pt}
Three flavor dynamical QCD project by CP-PACS/JLQCD
\thanks{Talk presented at 
        the 2nd Workshop on Lattice Hadron Physics (LHP2003),
        July 22-30, 2003, Cairns, Australia.}
}

\author{Kazuyuki Kanaya\address{Institute of Physics, 
        University of Tsukuba, Tsukuba, Ibaraki 305-8571, Japan}
        for CP-PACS and JLQCD Collaborations\thanks{Current members 
        of the CP-PACS and JLQCD Collaborations are:
        S. Aoki, M. Fukugita, S. Hashimoto, K. Ide, K.-I. Ishikawa, 
        T. Ishikawa, N. Ishizuka,
        Y. Iwasaki, KK, T. Kaneko, Y. Kuramashi, Y. Namekawa, 
        M. Okawa, T. Onogi, Y. Taniguchi, N. Tsutsui, A. Ukawa, N. Yamada, 
        T. Yamazaki, and T. Yoshi\'e.
        }
       }
\begin{document}

\begin{abstract}
The CP-PACS and JLQCD Collaborations have been made systematic studies
of lattice QCD carrying out both chiral and continuum extrapolations. 
Importance of dynamical quark effects has been clarified by a comparison 
of quenched QCD and two flavor full QCD simulations.
In two flavor simulations, the dynamical effects of $u$ and $d$ quarks are 
taken into account, but the third quark $s$ is still treated in a quenched 
approximation. 
As the final step towards a fully realistic lattice simulation of QCD, 
two collaborations started a joint project of three flavor QCD, 
concentrating all big computers available. 
Based on a series of preparative studies of exact simulation algorithm and 
non-perturbative improvement coefficient, a large scale simulation of 
three flavor QCD has been started.
I present the results of light hadron spectrum and light quark masses 
from the first production runs. 
\vspace{1pc}
\end{abstract}

\maketitle

\section{INTRODUCTION}

A direct calculation of hadronic properties from QCD is a fundamental 
objective in particle physics. Strong correlations among quarks make a 
reliable analytic calculation difficult even for basic properties of 
hadrons such as the mass spectrum and decay constants. 
So far numerical simulation based on the lattice formulation of QCD 
is the only reliable way toward this goal. 

A systematic study of lattice QCD carrying out both the chiral and 
continuum extrapolations requires a huge amount of computations, however. 
The CP-PACS and JLQCD Collaborations have been performing a series of large
scale simulations of QCD adopting the fastest computers available.
The CP-PACS Collaboration have been mainly using the CP-PACS computer.
a dedicated parallel computer with the peak performance of 614 GFLOPS.
The CP-PACS was designed and developed in 1996 at the Center for Computational 
Physics, University of Tsukuba \cite{CPPACScomputer}.
The main engine for the JLQCD Collaboration is the supercomputers at the 
High Energy Accelerator Research Organization (KEK). 
The central computer since 2000 is the Hitachi SR-8000/F1 
with 100 nodes achieving the peak speed of 1.2 TFLOPS \cite{KEKsc}.

As the first step, the CP-PACS Collaboration carried out an extensive
study of QCD in the quenched approximation, in which the effects of 
dynamical quark pair creations and annihilations are suppressed 
\cite{CPPACSquench}.
Performing the first well-controlled chiral and continuum extrapolations on 
lattices with the spatial extent of about 3 fm,
the limitation of the quenched approximation was clearly proven:
the light hadron mass spectrum deviates from the experiment by O(10\%).

The next natural step is to incorporate the effects of dynamical $u$, $d$ 
quarks (two flavor full QCD).
The results of systematic studies by the CP-PACS and JLQCD Collaborations
\cite{CPPACSfull2,JLQCDfull2} show that the discrepancies in the hadron mass 
spectrum observed in the quenched study are largely reduced by the dynamical 
$u$, $d$ quarks.
This demonstrates the importance of dynamical quarks.

In two flavor simulations, however, the third $s$ quark is treated 
in a quenched approximation yet. 
On the other hand, we do expect that contribution of dynamical $s$ quark is
not small because its mass is of the same order of magnitude as a typical 
energy scale of gluon dynamics: $m_s$ \lsim $\;\Lambda_{QCD}$.
Therefore, introduction of the dynamical $s$ quark in the simulation is 
the last major step left towards a fully realistic simulation of QCD.

\begin{table*}[t]
\begin{center}
\caption{
Computers for the three flavor QCD project by CP-PACS/JLQCD Collaborations. 
Performance of our PHMC code was measured in an actual production runs 
of three-flavor QCD on $20^3\times40$ lattices.}
\begin{tabular}{cccccc}
\hline
machine & location & \#nodes & peak speed & fraction for   & performance \\
        &          &        & [GFLOPS]   & LQCD [GFLOPS]  & of PHMC code \\
\hline
CP-PACS \protect\cite{CPPACScomputer} & CCP, U.Tsukuba 
         & 2048 & 614 & $\sim$614 & 20\% \\
SR-8000/G1 & CCP, U.Tsukuba & 12 & 173 & $\sim$173 & 44\% \\
SR-8000/F1 \protect\cite{KEKsc} & KEK & 100 & 1200 & $\sim$768 & 35\% \\
VPP-5000 \protect\cite{SIPC} & SICP, U.Tsukuba 
         & 80 & 768 & $\sim$230 & 44\% \\
Earth Simulator \protect\cite{ESC} & ES Center
         & 640 & 40960 & $\sim$640 & 31\% \\
\hline
\end{tabular}
\label{table:computers}
\end{center}
\vspace*{-3mm}
\end{table*}

The CP-PACS and JLQCD Collaborations have started a systematic study of 
three flavor full QCD. 
The first target is to perform a precise measurement of the light 
hadron spectrum and light quark masses. 

Because the required computer power is enormous, we have decided to 
concentrate all big computers available to us for this joint project. 
In Table~\ref{table:computers}, I list the major computers we are devoting 
to this project.
Fractions of the peak performance which we may fully use for lattice QCD 
simulations are given for each computer.
Summing up these numbers, we may use about 2.5 TFLOPS to simulate three
flavor QCD.

In this paper, I present the outline and status of the project.
The first step was to develop and test an exact algorithm for three-flavor 
QCD simulations. 
From a comparison of various variants of exact algorithms 
as described in Sec.~\ref{sec:PHMC}, 
we have chosen a polynomial hybrid Monte Carlo (PHMC) algorithm. 
In Sec.~\ref{sec:action}, I discuss our test study of three flavor QCD 
using the PHMC algorithm, which indicates that improvement of lattice action 
is essential for meaningful simulations. 
We thus adopt the RG-improved gauge action by Iwasaki 
and the clover-improved Wilson quark action. 
To achieve full $O(a)$-improvement, we then determined the non-perturbative 
value of the clover coefficient $c_{SW}$. 
Based on these preparative studies, we have now started a large scale 
simulation of three flavor QCD.
Results of the first production runs are presented in Sec.~\ref{sec:run}.
Tentative conclusions are given in Sec.~\ref{sec:conclusion}.

\begin{figure}[t]
\centerline{
a)\epsfxsize=7.2cm\epsfbox{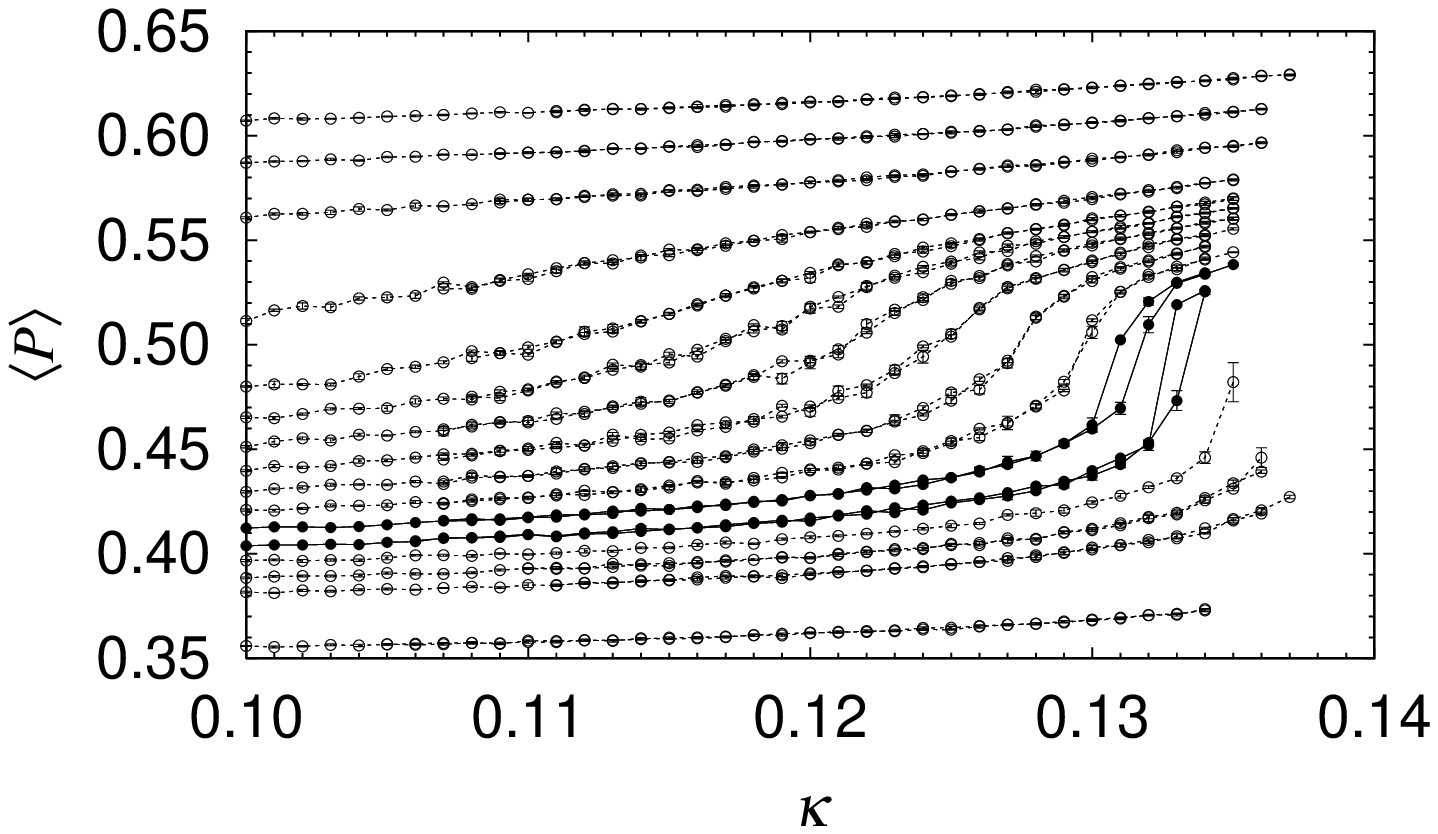}
}
\vspace{1mm} 
\centerline{
b)\epsfxsize=7.2cm\epsfbox{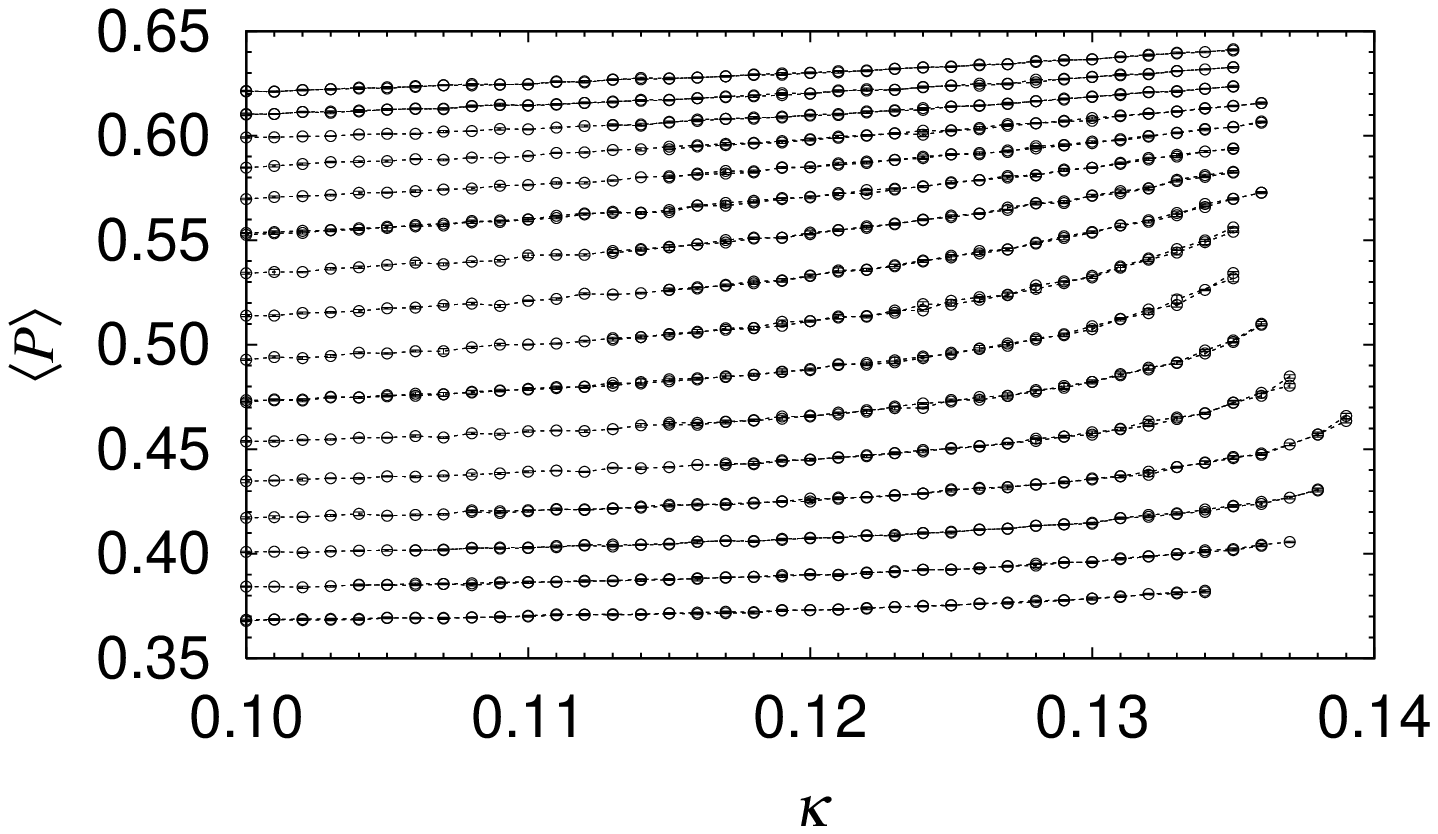}
}
\vspace*{-8mm}
\caption{Thermal cycles for plaquette obtained on $8^3\times16$ 
lattices at $a^{-1}\!\simeq\!1.5$--2 GeV \protect\cite{OkawaLat01}.
a) Plaquette gauge action at $\beta=4.6$, 4.8, $\cdots$, 6.0 from bottom
to top.
b) RG gauge action at $\beta=1.5$--2.25 in steps of 0.05 from bottom to top.
}
\label{fig:test}
\vspace*{-2mm}
\end{figure}

\section{EXACT ALGORITHM}
\label{sec:PHMC}

With the Wilson-type (staggered-type) lattice quarks, 
the exact HMC algorithm exists only for the cases of even 
(four times integer) number of flavors. 
Because an exact algorithm for odd number of flavors was not known
before, inexact R-algorithm has been adopted in previous studies of 
three flavor QCD.
To avoid possible unexpected systematic errors, however, it is highly 
desirable to adopt an exact algorithm for large scale simulations.

Recently, several exact algorithms for odd flavors have 
been proposed: the multi-boson algorithm \cite{Alexandrou99} and 
the polynomial hybrid Monte Carlo (PHMC) algorithm \cite{PHMCalgo1,PHMCalgo2}, 
both based on polynomial approximation for the quark matrix and for its 
determinant.
We adopt the PHMC algorithm extending it to clover-improved Wilson quarks. 
We made a systematic test of several variants of the PHMC 
algorithm \cite{JLQCD_PHMC}.
We found that, with appropriate improvements, the PHMC algorithm is equally 
efficient as the HMC algorithm for the case of two-flavor QCD.

From this study, we decide to adopt the PHMC algorithm for the $s$ quark and 
the HMC algorithm for degenerate $u$, $d$ quarks. 
Note that, because the $s$ quark is heavy, the problem of negative determinant 
at very small quark masses is automatically avoided. 
Together with additional improvements, we find that 
the overall CPU time to simulate one trajectory of three flavor QCD is 
only about 1.5 times more than that for two flavor QCD using our previous 
HMC code at the same $u$, $d$ quark masses.
We conclude that the PHMC algorithm is sufficiently 
efficient to carry out a systematic simulation of three flavor QCD 
with the present power of computers.

We have implemented the code to the computers listed in 
Table~\ref{table:computers}.
Optimizing the vectorization and parallelization algorithms 
depending on the characteristics of each machine, we achieved 
the performance of 20--44\% in actual production runs \cite{YoshieLat03}, 
as compiled in the last column of Table~\ref{table:computers}.

\section{CHOICE OF THE LATTICE ACTION}
\label{sec:action}

Improvement of the lattice action is effective to suppress lattice 
artifacts on coarse lattices \cite{comparative} 
and has played an essential role in our study of two flavor QCD to 
reduce necessary computer resources.
Because the requirement of computer power is even more oppressive 
in the three flavor project due to larger number of parameters, 
improvement will be important for a systematic simulation of three 
flavor QCD.

\subsection{A test study}
Using the exact PHMC algorithm, we made a series of test studies 
at $a^{-1}\!\simeq\!1.5$--2 GeV on $8^3\times16$ and $12^3\times32$ 
lattices \cite{OkawaLat01}.
For gluons, we test the standard one-plaquette action, the RG-improved
action by Iwasaki \cite{Iwasaki}, and the meanfield-improved Symanzik 
gauge action.
For quarks, we adopt the clover-improved Wilson quark action with 
meanfield-improved clover-coefficient $c_{SW}$.

With the plaquette gauge action, we have encountered a severe lattice
artifact at $a^{-1}$ \lsim $\;2$ GeV:
the plaquette expectation values shown in Fig.~\ref{fig:test}(a) 
indicate unexpected first-order transitions at $\beta \! = \! 4.95$ and 5.0.
Our study of the lattice size dependence suggest that this is a bulk 
transition.
On the other hand, results with improved gauge actions show no signs
of hysteresis, as shown in Fig.~\ref{fig:test}(b) for the RG-improved 
action.
We suspect that the lattice artifact in the case of plaquette action is
due to an effective adjoint coupling from the clover term \cite{OkawaLat01}.

In any case, because it is difficult to simulate several points 
beyond $a^{-1}\!\sim\!2$ GeV, our findings imply that 
improvement of the gauge action is indispensable to perform a continuum 
extrapolation.
We adopt the RG-improved gauge action for gluons.

\subsection{Non-perturbative $C_{SW}$}

For quarks, we adopt clover-improved Wilson quark action.
To completely remove $O(a)$ errors, we need non-perturbative values of 
the clover coefficient $c_{SW}$, which have not been estimated for three
flavor QCD.
In \cite{IshikawaLat02,IshikawaLat03}, we have determined 
the non-perturbative $c_{SW}$ 
both for the plaquette and RG-improved gauge actions.
We use the Schr\"odinger functional method \cite{SF}. 
For the RG gauge action we adopt the boundary condition 
described in \cite{SFRG}.

We found that the finite volume effect in $c_{SW}$ is not negligible for 
the case of the RG-improved gauge action. 
When we estimate $c_{SW}$ at a fixed dimension-less lattice size $L/a$ 
as usually done in previous studies, 
$c_{SW}$ suffers a constant deviation from the true non-perturbative value
even in the continuum limit.
This yields $O(a)$ errors in physical observables. 

\begin{figure}[t]
\vspace*{-3mm}
\centerline{
\epsfxsize=7.7cm\epsfbox{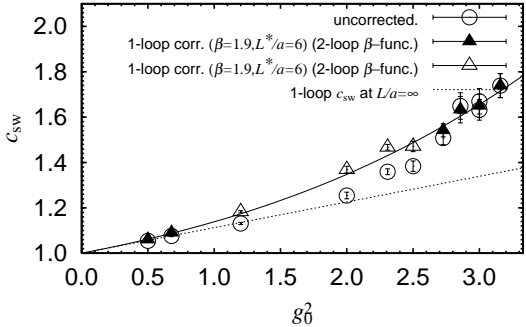}
}
\vspace*{-9mm}
\caption{Non-perturbative value of $c_{SW}$ in three-flavor QCD with 
RG-improved gauge action \protect\cite{IshikawaLat03}.
Open and filled triangles show $c_{SW}$ at a fixed physical lattice size 
$L^*$, obtained by correcting the raw results shown by open circles.
Dotted line shows the one-loop result for $L/a=\infty$.
}
\label{fig:Csw}
\vspace*{-2mm}
\end{figure}

In order to remove the finite volume effect, we estimate the non-perturbative 
$c_{SW}$ at a fixed physical lattice size $L^*$. 
Then the finite volume effect vanishes as $a/L^*$ in the continuum limit.
Correction of the data to $L^*$ from the results obtained at 
the simulation point $L$ is done using one-loop formulae calculated 
with the SF setup in finite volume \cite{SFRG}. 
See \cite{IshikawaLat03} for details.  

Our final results for $c_{SW}$ at $L^* = 6 a_{\beta=6/g^2_0 = 1.9}$ are 
summarized in Fig.~\ref{fig:Csw} by triangles. 
We note that the results smoothly converge to the one-loop result for 
$L/a=\infty$ (dashed line) in the continuum limit $g_0=0$.

\section{THREE FLAVOR QCD}
\label{sec:run}

\begin{figure}[t]
\centerline{
\epsfxsize=6.5cm\epsfbox{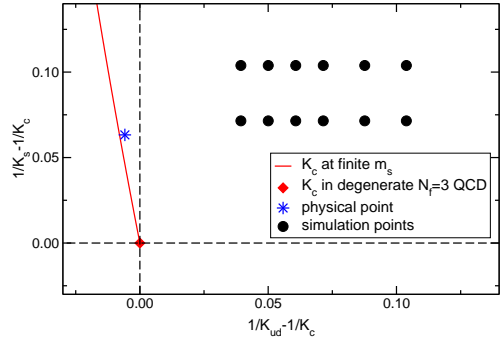}
}
\vspace*{-8mm}
\caption{Simulation points on $16^3\times32$ lattice at $\beta=1.9$.
Our estimation for the physical point is given by the star.
The line shows the chiral limit defined by $m_{\rm PS,LL}=0$.
}
\label{fig:Kcp}
\vspace*{-2mm}
\end{figure}

Based on the preparative studies discussed in previous sections, we have 
started a systematic simulation of three flavor QCD. 
As the first point towards the continuum limit, 
we are carrying out jobs at $a^{-1}\!\simeq\!2$ GeV ($a\!\simeq\!0.1$ fm).
We have finished the production runs on a $16^3\times32$ lattice
($La\!\sim\!1.6$ fm), and are now carrying out simulations on a 
$20^3\times40$ lattice ($La\!\sim\!2.0$ fm).
In this section, I present the results of the first production runs on 
the $16^3\times32$ lattice \cite{KanekoLat03}.

\subsection{Simulation parameters}

With the RG-improved gauge action and clover-improved Wilson quark action,
we made simulations at $\beta=1.9$ on a $16^3\times32$ lattice 
\cite{KanekoLat03}.
The non-perturbative value of $c_{SW}$ at this $\beta$ is 1.715 
\cite{IshikawaLat03}.
We studied six values of the $u,d$ quark mass in the range 
$K_{ud} \! = \! 0.1358$--0.1370 corresponding to 
$m_{\rm PS,LL}/m_{\rm V,LL} \! \simeq \! 0.64$--0.77,
where L means the light $u,d$ sea quark.
For the $s$ quark mass, we studied two points 
$K_s \! = \!0.1364$ and 0.1358 corresponding to 
$m_{\rm PS,SS}/m_{\rm V,SS} \! \simeq \! 0.72$ and 0.77,
where S is for the sea $s$ quark. 
These values are close to the physical $s$ quark point 
$m_{\eta_s}/m_{\phi} \! \simeq \! 0.68$ from the chiral perturbation theory.
The simulation points are summarized in Fig.~\ref{fig:Kcp}.

We simulated 3000 trajectories at each $(K_{ud},K_s)$ 
and accumulated the configurations every 10 trajectories. 
The HMC step size and the order of polynomial in the PHMC algorithm 
were adjusted to achieve the acceptance rate of more than 85\% 
in HMC steps and more than 90\% in PHMC steps.
So far, we have measured hadronic observables only when each 
valence quark is one of the sea quarks, L or S.
Errors are estimated by a jack-knife method with bins of 50 trajectories.

\begin{figure}[t]
\centerline{
\epsfxsize=6.5cm\epsfbox{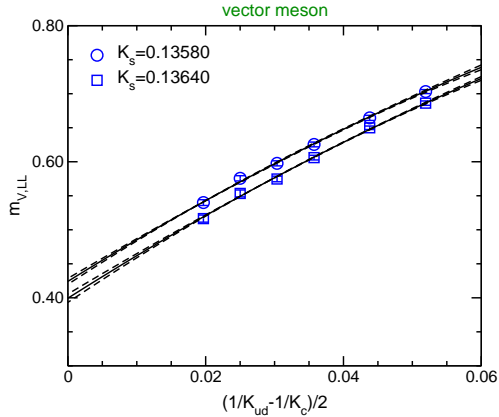}
}
\vspace*{-8mm}
\caption{
Vector meson mass $m_{\rm V,LL}$ at $\beta=1.9$ on $16^3\times32$ lattice. 
}
\label{fig:mV_Kinv}
\vspace*{-2mm}
\end{figure}

\begin{figure}[t]
\centerline{
\epsfxsize=7.5cm\epsfbox{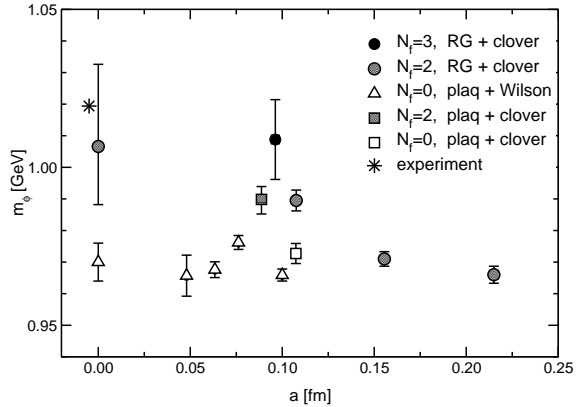}
}
\vspace*{-7mm}
\caption{
$\phi$ meson mass from the $K$-input as a function of the lattice spacing. 
The result from three flavor QCD is given by the filled circle at 
$a \! \simeq \! 0.1$ fm.
Quenched (open symbols) \cite{CPPACSquench,CPPACSfull2}
and two flavor QCD results (shaded symbols) \cite{CPPACSfull2,JLQCDfull2}
are also shown.
}
\label{fig:mphi}
\vspace*{-2mm}
\end{figure}

\begin{figure}[t]
\centerline{
\epsfxsize=7.2cm\epsfbox{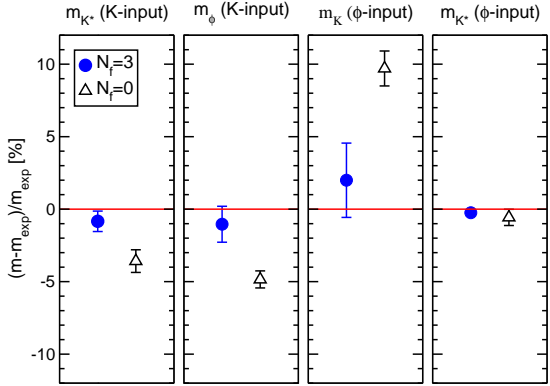}
}
\vspace*{-6mm}
\caption{
Relative discrepancy of meson masses in three flavor QCD 
at $a^{-1}\!\simeq\!2$ GeV \protect\cite{KanekoLat03}. 
Results of quenched QCD in the continuum limit
are also shown by open symbols.
}
\label{fig:disc}
\vspace*{-2mm}
\end{figure}

\subsection{Meson spectrum}

Because the spatial lattice size of $La \sim 1.6$ fm is not quite large, 
we concentrate on mesons in this report.
We tested point and exponentially smeared sources.
Because we obtained clearest plateaus for effective masses when 
both quark sources are smeared, we show the results with doubly 
smeared sources in the following.

Results for the mass of vector mesons consisting of two light sea quarks 
are shown in Fig.~\ref{fig:mV_Kinv} as a function of $1/K-1/K_c$,
where $K_c$ is defined by $m_{\rm PS,LL}(K_{ud}\!=\!K_s\!=\!K_c) = 0$.
Results for other mesons are similar.
Because the quark mass dependence is quite smooth, we adopt following 
polynomial ans\"atze
\begin{eqnarray}
m_{\rm PS}^2 &=& B m_{q,sea} + 
\nonumber \\
&& (C + D m_{q,sea}) (m_{q,val1}+m_{q,val2})
\label{eq:ps}
\\
m_{\rm V} &=& A' + B' m_{q,sea} + 
\nonumber \\
&& (C' + D' m_{q,sea}) (m_{q,val1}+m_{q,val2})
\label{eq:v1}
\end{eqnarray}
or
\begin{eqnarray}
m_{\rm V} &=& A'' + B'' \mu_{sea} + 
\nonumber \\
&& (C' + D' \mu_{sea}) (m_{\rm PS,11}^2 + m_{\rm PS,22}^2)
\label{eq:v2}
\end{eqnarray}
where
$m_{q,sea} = 2 m_{ud} + m_s$ with $m_q=(1/K_q - 1/K_c)/2$ 
and $\mu_{sea} = 2 m_{\rm PS,LL}^2 + m_{\rm PS,SS}^2$. 
We test both combinations of (\ref{eq:ps})$+$(\ref{eq:v1}) and 
(\ref{eq:ps})$+$(\ref{eq:v2}). 
Both ans\"atze fit the data well.
We quote the weighted average of two combinations as the central value of 
the masses, while the difference between two combinations is treated 
as a systematic error from the chiral fit.
To identify the lattice spacing $a$ and the physical point 
$(K_{ud}^{phys},K_s^{phys})$, we use either $(M_\pi,M_\rho,M_K)$ ($K$-input)
or $(M_\pi,M_\rho,M_\phi)$ ($\phi$-input).

Result for $M_\phi$ from the $K$-input is shown in Fig.~\ref{fig:mphi}.
The clear discrepancy between experiment (star) and quenched results 
(open symbols) is largely removed in two flavor QCD (shaded symbols) 
in the continuum limit.
Our new three flavor value is shown by the filled symbol at 
$a \simeq 0.1$ fm.
We find that the three flavor result is higher than the two flavor results 
and, already at $a \simeq 0.1$ fm, the mass is 
consistent with experiment.
Result for $M_{K^*}$ given in \cite{KanekoLat03} is similar.

We summarize the relative discrepancy with experiment 
in Fig.~\ref{fig:disc}.
We see that experimental spectrum is well reproduced in three flavor QCD
at $a \simeq 0.1$ fm.
Reproduction of experimental values can be confirmed also by the 
$J$ parameter \cite{KanekoLat03} for which a chiral fit is not required.
Consistency with experiment at finite $a$ imply that the $K$- and 
$\phi$-inputs lead to consistent values already at finite $a$.
Accordingly, we obtain $a^{-1} = 2.05(4)$ GeV from the $K$-input and 
2.05(5) GeV from the $\phi$-input.

For a precise prediction, we need to extrapolate the results to the 
continuum limit.
However, we may hope that the scaling violation is indeed small 
for our choice of non-perturbatively $O(a)$-improved action, 
such that the consistency with experiment is kept in the continuum limit.
This should be tested in future.

\begin{figure}[t]
\centerline{
a) \epsfxsize=7.2cm\epsfbox{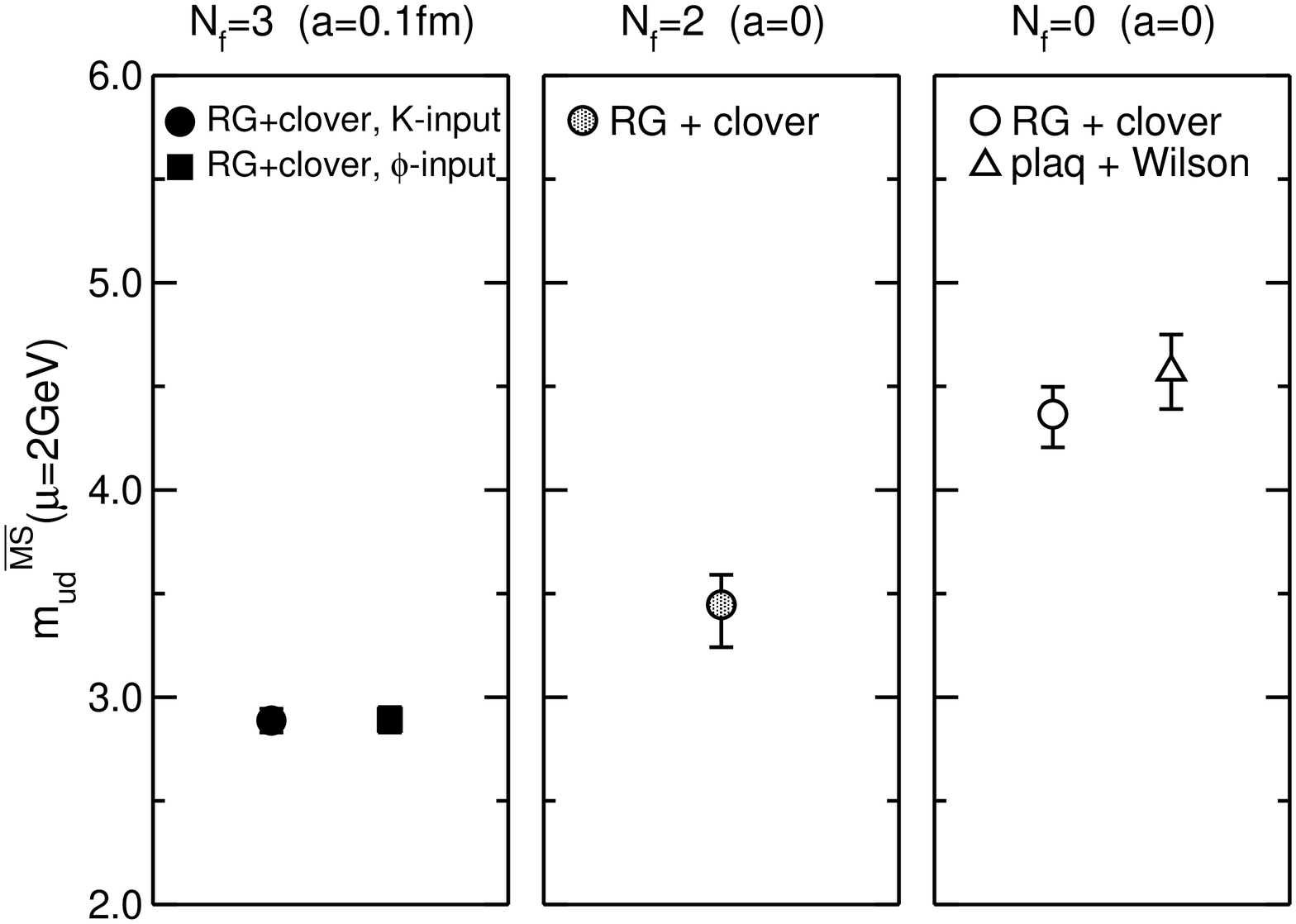}
}
\vspace{5mm} 
\centerline{
b) \epsfxsize=7.2cm\epsfbox{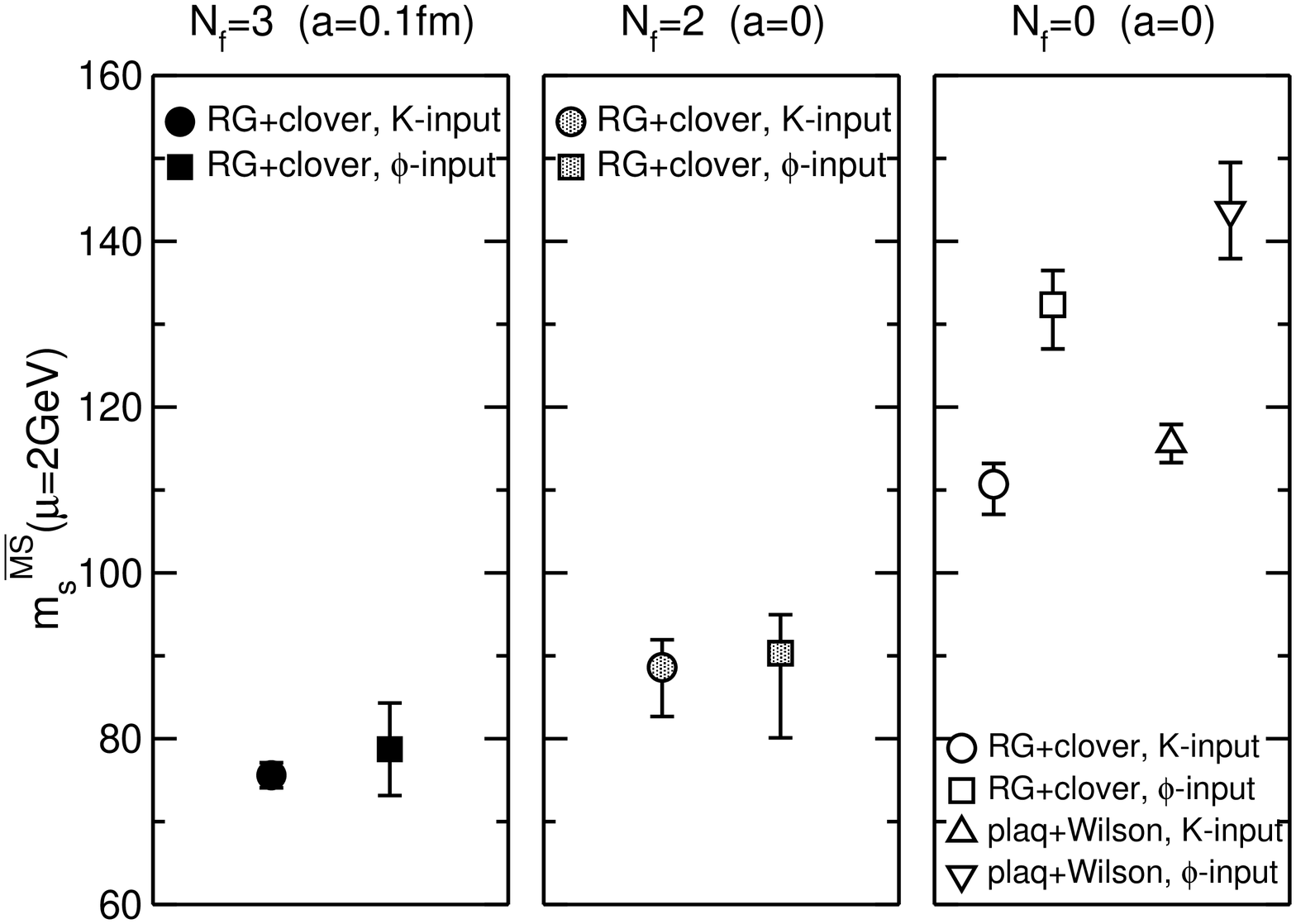}
}
\vspace*{-6mm}
\caption{
Quark masses (a) $m_{ud}$ and (b) $m_s$ in the $\overline{\rm MS}$ scheme 
at $\mu\!=\!2$~GeV \protect\cite{KanekoLat03}.
}
\label{fig:mq}
\vspace*{-2mm}
\end{figure}

\subsection{Quark mass}

Because quarks are confined, their masses are not direct observables of the 
theory. 
Therefore, there exist several alternative definitions for the quark mass. 
Two popular definitions are the axial-vector Ward identity (AWI) quark mass
\begin{equation}
m_q = Z_q \frac{\left\langle \Delta_4 A_4 P\right\rangle}
              {2 \left\langle P P\right\rangle}
\label{eq:AWImq}
\end{equation}
and the vector Ward identity (VWI) quark mass 
\begin{equation}
m_q = Z_q' (1/K - 1/K_c)/2,
\label{eq:VWImq}
\end{equation}
where $A_4$ is the fourth component of the axial-vector current, 
$P$ is the pseudo-scalar density, and $Z$'s are renormalization factors.
Different definitions lead to different values for $m_q$ on finite lattices. 
This was a big source of error in early calculations of $m_q$.
In our previous studies of quenched and two flavor full QCD, however, 
we have shown that they converge to universal values in the continuum limit 
\cite{CPPACSquench,CPPACSfull2}.

In three flavor QCD, we found that, 
although the differences between $K$- and $\phi$-inputs are absent, 
the AWI and VWI quark masses disagree at $a^{-1}\!\simeq\!2$ GeV. 
We also noted that the choice of ``$K_c$'' in (\ref{eq:VWImq}) introduces a
sizable ambiguity in the values of VWI quark mass: 
From Fig.~\ref{fig:Kcp}, we obtain even negative value of $m_{ud}$ when we 
adopt $K_c$ defined by $m_{\rm PS,LL}(K_{ud}\!=\!K_s\!=\!K_c) = 0$.
Such ambiguities will be removed in the continuum limit. 

At the present stage having data only at a value of $a^{-1}$, we would like to 
concentrate on the AWI quark mass which does not suffer from the ambiguity 
of $K_c$ and, in the case of two flavor QCD, shows a smaller scaling violation 
than VWI quark masses \cite{CPPACSfull2}.
To convert to the quark mass in the $\overline{\rm MS}$ scheme, we 
match the renormalized mass with the lattice data at $\mu\!=\!a^{-1}$ 
using a meanfield-improved one-loop $Z$ factor \cite{matching}, 
and let it run to $\mu\!=\!2$~GeV 
using the 4-loop beta function. % \cite{4loop-running}.

Our results are summarized in Fig.~\ref{fig:mq}.
Previous results from quenched and two flavor full QCD, extrapolated to 
the continuum limit, are also shown for comparison 
\cite{CPPACSquench,CPPACSfull2}.
We see that quark masses decrease when we increase the number of 
dynamical quark flavors. 
In three flavor QCD, we obtain 
$m_{ud}^{\overline{\rm MS}} = 2.89(6)$ MeV and
$m_{s}^{\overline{\rm MS}} = 75.6(3.4)$ MeV
at $a^{-1}\!\simeq\!2$ GeV, 
where the central values are from the $K$-input and the errors include 
systematic errors estimated by the difference between 
the $K$- and $\phi$-inputs.
With the effects of the dynamical $s$ quark, 
both $m_{ud}$ and $m_s$ are decreased by about 15\% from the previous
two flavor values.
The ratio $m_{s}/m_{ud} = 26.2(1.0)$ is 
consistent with the one-loop estimate of 
chiral perturbation theory 24.4(1.5) \cite{mq.ChPT}.

\section{TENTATIVE CONCLUSIONS AND OUTLOOK}
\label{sec:conclusion}

We have presented the status of the joint project of three flavor QCD
simulation by the CP-PACS and JLQCD Collaborations.
We employ exact HMC algorithm for light $u,d$ quarks and exact PHMC 
algorithm for $s$ quark. 
To reduce the lattice artifacts, we adopt the RG-improved gauge action 
and non-perturbatively $O(a)$-improved clover quark action.
As the first step toward a systematic study of the fully realistic QCD on 
the lattice, we made simulations at $a^{-1} \! \simeq \! 2$ GeV
on $16^3\times 32$ lattices. 
These simulations show that, with the effects of dynamical $s$ quark, 
light meson mass spectrum agrees well with experiment 
already at $a^{-1} \! \simeq \! 2$ GeV.
We also found that the quark masses $m_{ud}$ and $m_s$ are lower than the 
two flavor values by about 15\%.
We are currently performing simulations at the same simulation points 
on $20^3\times40$ lattices to study finite lattice volume effects.
Although a continuum extrapolation is not made yet, these results are 
quite encouraging to further carry out large scale simulations of 
three flavor QCD.

\vspace*{3mm}
I thank the members of the CP-PACS/JLQCD Collaborations for discussions.
This work is in part supported 
by the Large Scale Simulation Program No. 98 (FY2003) of High Energy 
Accelerator Research Organization (KEK), 
by Large Scale Numerical Simulation Project of S.I.P.C., Univ. of Tsukuba,
by Earth Simulator Research Projects, ES Center,
and by the Grants-in-Aid of the Ministry of Education
(Nos. 1315204 and 13640260).

%%%%%%%%%%%%%%%%%%%%%%%%%%%%%%%%%%%%%%%%%%%%%%%%%%%%%%%%%%%%%

\end{document}